\documentclass[3p,times,twocolumn]{elsarticle}

\usepackage{amsmath}

\newcommand{\GeV}{\,{\rm GeV}}

\newcommand{\matel}[3]{\langle #1|#2|#3\rangle}
\newcommand{\mi}{\!-\!}
\newcommand{\C}{\mathbb{C}}

\usepackage{amssymb}

\usepackage[figuresright]{rotating}

\begin{document}
\begin{frontmatter}




\title{Heavy-to-light  chromomagentic matrix element}

\author[label1]{Maria Dimou}
\author[label1,label2]{James Lyon}
\author[label1,label2,label3]{Roman Zwicky}

\address[label1]{School of Physics \& Astronomy, University of Southampton, 
    Highfield, Southampton SO17 1BJ, UK}
  
\address[label2]{School of Physics \& Astronomy, University of Edinburgh, 
    Edinburgh EH9 3JZ, Scotland}
 \address[label3]{speaker: (CP3-Origins-2013-016 DNRF90,DIAS-2013-16)}

\begin{abstract}
We report the computation of the matrix element of the chromomagnetic operator 
of the flavour changing neutral current (FCNC)-type 
between a $B$- or $D$-meson state and a light hadron and off-shell photon. 
The computation is carried out by using  the method of light-cone sum rules (LCSR).  It is found that the
matrix element exhibits a large strong phase for which we give a long distance interpretation.
The analytic structure of the correlation function in use admits a complex anomalous threshold 
 on  the physical sheet, the meaning and handling of which within the sum rule approach 
is discussed. We  compare our results to QCD factorisation for which spectator photon 
emission is end-point divergent.
\end{abstract}







\end{frontmatter}


\section{Introduction}
\label{}
The chromomagnetic operator of the $b \to s$-type is defined as follows:
\begin{equation}
\label{eq:O8}
\tilde  {\cal O}_8^{(')}   =    \bar s \sigma_{\mu\nu}  G_a^{\mu\nu}\frac{\lambda^a}{2}  (1\pm \gamma_5) b \;.
 \end{equation}
In the   effective Hamiltonian\footnote{The coefficient $k$ is convention dependent and $C_8$ is the Wilson coefficient.} ${\cal H}^{\rm eff}_{b \to s} = 
  \, k G_F C_8 \,  [ m_b(m_s) \tilde  {\cal O}_8] + ..$, 
 there is an additional factor of $m_b(m_s)$, whose origin can be understood from the minimal flavour 
 symmetry of the Standard Model (SM).  The operator is therefore effectively of mass dimension six 
and thus on the same footing as the four-Fermi interactions.

The motivation for our work \cite{DLZ12}  is twofold:
first, to provide an estimate of the chromomagnetic matrix element; and second, to compare it with the QCD factorisation (QCDF) calculation.
The matrix element is of importance for isospin asymmetries  \cite{KN01,FM02,pQCD,BJZ,LZ12b},
as the photon can be emitted from the spectator quark, and for testing for  weak phases
in $C_8^{(')}$ \cite{IK12,LZ12a}, which may be related to the relatively large direct CP-violation 
in $D^0 \to \pi^+\pi^-/K^+K^-$  \cite{LHCbCHARM}.
The comparison with QCDF is not natural as LCSR are
not tailored around the heavy quark expansion and contain additional contributions not inherent in the QCDF calculation.
Possibly the most surprising outcome of our investigations 
is the appearance of a complex anomalous threshold on the physical sheet in the correlation function used 
to extract the matrix element.

\section{Definitions and computation}
\label{sec:defSR}
We aim to compute the following matrix element
\begin{eqnarray}
\label{eq:matel}
 {\cal A}^{* \rho}(V)   &=& \matel{\gamma^*(q,\rho) V(p,\eta)}{\tilde O_8}{\bar B(p_B)}  \;,
\end{eqnarray}
where $V$ stands for a light vector meson of the $\rho, K^*$,etc.-type 
and the star indicates that the photon $\gamma$ can be off-shell, i.e. $q^2 \neq 0$. 
The formalism allows us to extract the matrix element above
with the $B$-meson replaced by a $D$-meson as well as the $V$-meson replaced
by a light pseudoscalar of the $\pi, K$,etc.-type.  
Throughout this write-up we shall refer mostly to the  $\bar B \to V \gamma^*$ transition and
replacements for the other decays are considered implied.
Eq.~(\ref{eq:matel}) decomposes into the following transverse Lorentz structures\footnote{The factor $c_V$ is inserted to absorb trivial factors
due to the $\omega \sim (\bar u u + \bar d d)/\sqrt{2} , \rho^0 \sim (\bar u u - \bar d d)/\sqrt{2}  $ wave functions. $c_V= - \sqrt{2}$ for $\rho^0$ in $b \to d$, $c_V = \sqrt{2}$ in 
all other transitions into $\omega$ and $\rho^0$, and $c_V=1$ otherwise. }
\begin{eqnarray}
\label{eq:FFdec}
c_V \, {\cal A}^{*\rho}(V)  &=&
 k_G \, \sum_{i=1}^3G_i(q^2) P_i^\rho   \;,
 \nonumber  \\[0.2cm]
 {\cal A}^{*\rho}(P)   &=&
k_G \,  G_T (q^2) P_T^\rho   \;.
 \end{eqnarray}
($q_\rho P_\iota^\rho = 0$ where $\iota \in \{1,2,3,T\}$)
\begin{align}
\label{eq:Vprojectors}
P_1^\rho =&  2 \epsilon^{\rho}_{\phantom{x} \alpha \beta \gamma} \eta^{*\alpha} p^{\beta}q^\gamma  \;, \nonumber \\
P_2^\rho =& i \{(m_B^2\mi m_V^2) \eta^{*\rho} \mi 
(\eta^*\!\cdot\! q)(p+p_B)^\rho\}  \;, \nonumber \\
P_3^\rho =&  i(\eta^*\!\cdot\! q)\{q^\rho \mi  \frac{q^2}{m_B^2\mi m_V^2} (p+p_B)^\rho \}   \;,  \nonumber \\
P_T^\rho  =&  \frac{1}{m_B+m_K}  \{(m_B^2-m_K^2)q^\rho \mi q^2 (p+p_B)^\rho\}  \;.
\end{align}
The prefactor $k_G \equiv  -2 e/g  $ is chosen such that $G_i$ and $G_T$ parallel 
the standard form factors $T_1$ and $f_T$ in the sense that the amplitude reads
${\cal A}(b \to s) \propto (C_7 T_1 + C_8 G_1)P_1 + .. $ and likewise in the pseudoscalar case. 
Under the replacement ${\cal O}_8 \to {\cal O}'_8$, i.e. $(1+\gamma_5) \to (1-\gamma_5)$, at our level of approximation\footnote{The sign alternate from chiral even to chiral odd DA.},  
the matrix element transforms as follows,
\begin{equation}
\label{eq:prime}
\{G_1,G_2,G_2,G_T\}  \; \stackrel{\gamma_5 \to -\!\gamma_5}{ \to} \;  \{G_1,-G_2,-G_3,G_T\} \;,
\end{equation}
by virtue of parity conservation of QCD.

\begin{figure}[h]
 \centerline{\includegraphics[width=3.6in]{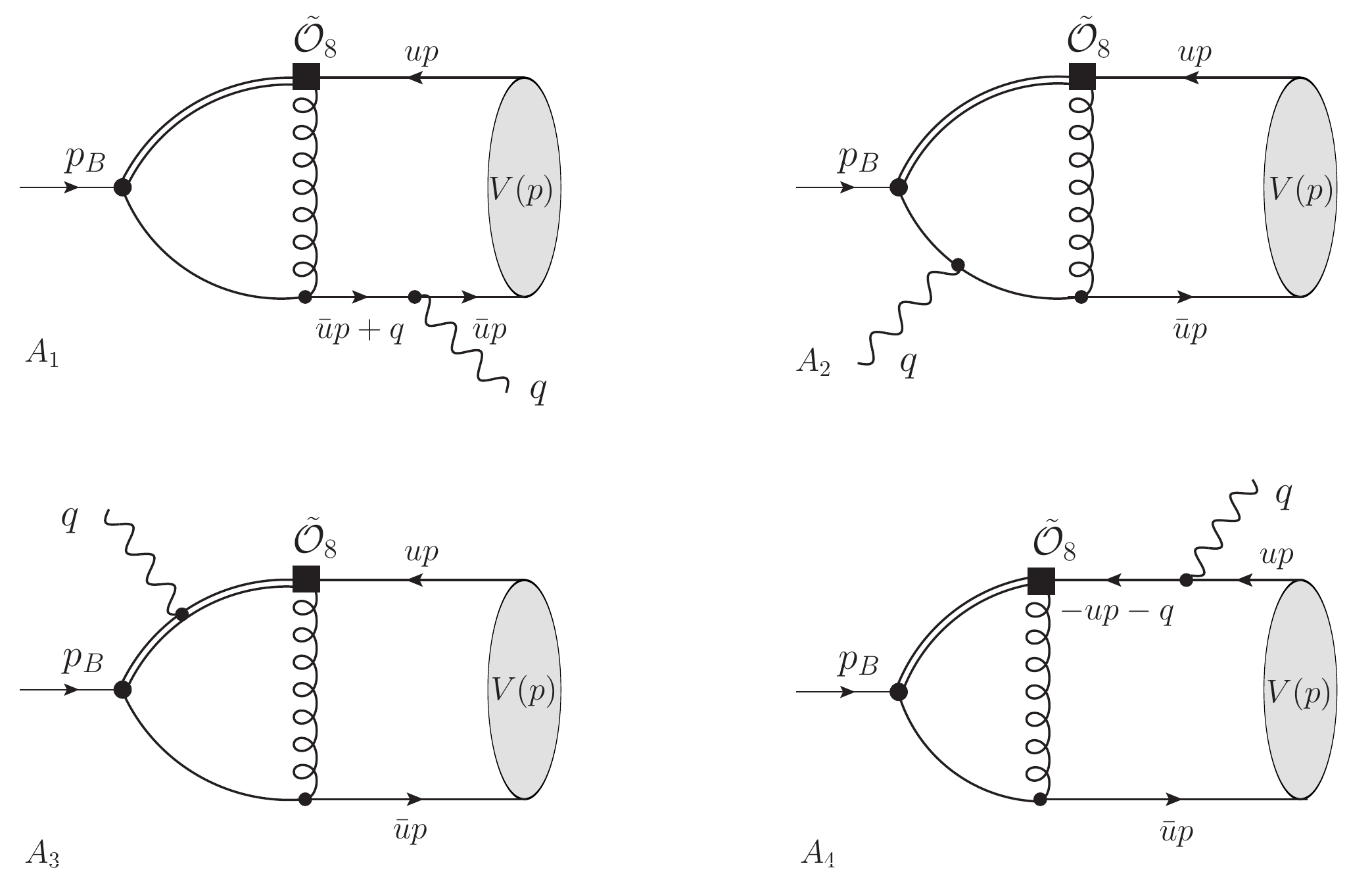}}
  \centerline{
 \includegraphics[width=3.6in]{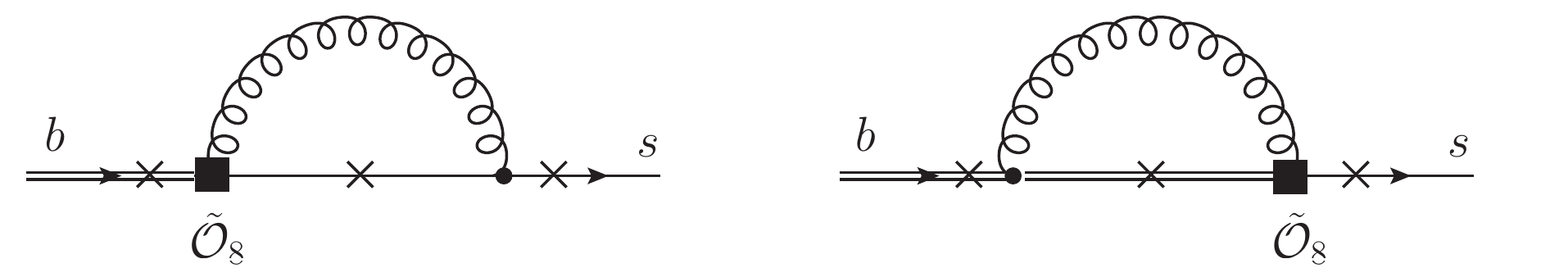}
}
 \caption{\small (top/middle) Diagrams $A_1$ to $A_4$ to which we refer to as spectator corrections 
 (bottom) Non-spectator corrections.
Crosses denote possible places for photon emission.}
 \label{fig:diaA}
 \end{figure}

The matrix element is extracted from the following correlation
function:\footnote{For the sake of notational simplicity, we shall keep the photon polarisation tensor contracted here as in (\ref{eq:matel}), though from a physical point of view this
does not make sense  for an off-shell photon.}
\begin{equation}
\label{eq:CF}
\!\!\! \Pi^V =   i \!\! \int_x \matel{\gamma^*(q) V(p) }{T J_B(x) \tilde{\cal O}_8(0)}{0} e^{-\! i p_B \cdot x} ,
\end{equation}
where $J_B  = i m_b \bar b \gamma_5 q$  plays the role of the interpolating current for the 
$B$-meson. 
At leading order in $\alpha_s$ there are a total of twelve graphs. 
We divide these into those where the gluon connects to the spectator quark~(s) and those where it connects to 
the non-spectator quark~(ns):
\begin{equation}
\label{eq:decsns}
G_\iota(q^2)  = G_\iota^{(s)}(q^2) + G_\iota^{(ns)}(q^2) \;.
\end{equation}
The four diagrams denoted by  $A_1$ to $A_4$  in Fig.\ref{fig:diaA}(top,middle) contribute to 
$G_\iota^{(s)}$ whereas the diagrams at the bottom of the same figure correspond to the 
$G_\iota^{(ns)}$-contributions. The latter  factorise into a function of $f(q^2/m_b)$ times
standard vector, axial or tensor form factors. The function $f$, in terms of an expansion in powers of 
$q^2/m_b^2$ and logarithmic terms, has been obtained in  
the inclusive case in  \cite{AAGW01}\footnote{We would like to add that it would be possible 
to  compute these contributions within LCSR.}. 
The two diagrams where the photon is emitted from the spectator quark
and the gluon connects to the non-spectator quark are not shown.  They are expected to be small
since no fraction of the $m_b$-rest mass is transmitted to the energetic photon.

The diagrams are computed via the light-cone operator product expansion (LC-OPE), since 
the correlation function is believed to be dominated by light-like distances.    Schematically this amounts to
to a convolution between a perturbatively calculable hard scattering kernel $T_H$ and light-cone distribution amplitudes (DA) $\phi$: 
$$
\Pi(q^2,p_B^2) =  \sum_i T^{(i)}_H(q^2,p_B^2;\mu_F;u) \circ \phi^{(i)}(u,\mu_F) \;.
$$
 The sum extends over increasing orders of twist, defined as
 the dimension of the operator minus its spin projection,
and the variable $u$ stands for the momentum fraction of the strange quark in the light meson.
We limit ourselves to leading twist-2.

\section{Sum rules and anomalous thresholds}
\label{sec:SRanomalous}

\begin{figure}[h]
 \centerline{ \includegraphics[width=3.4in]{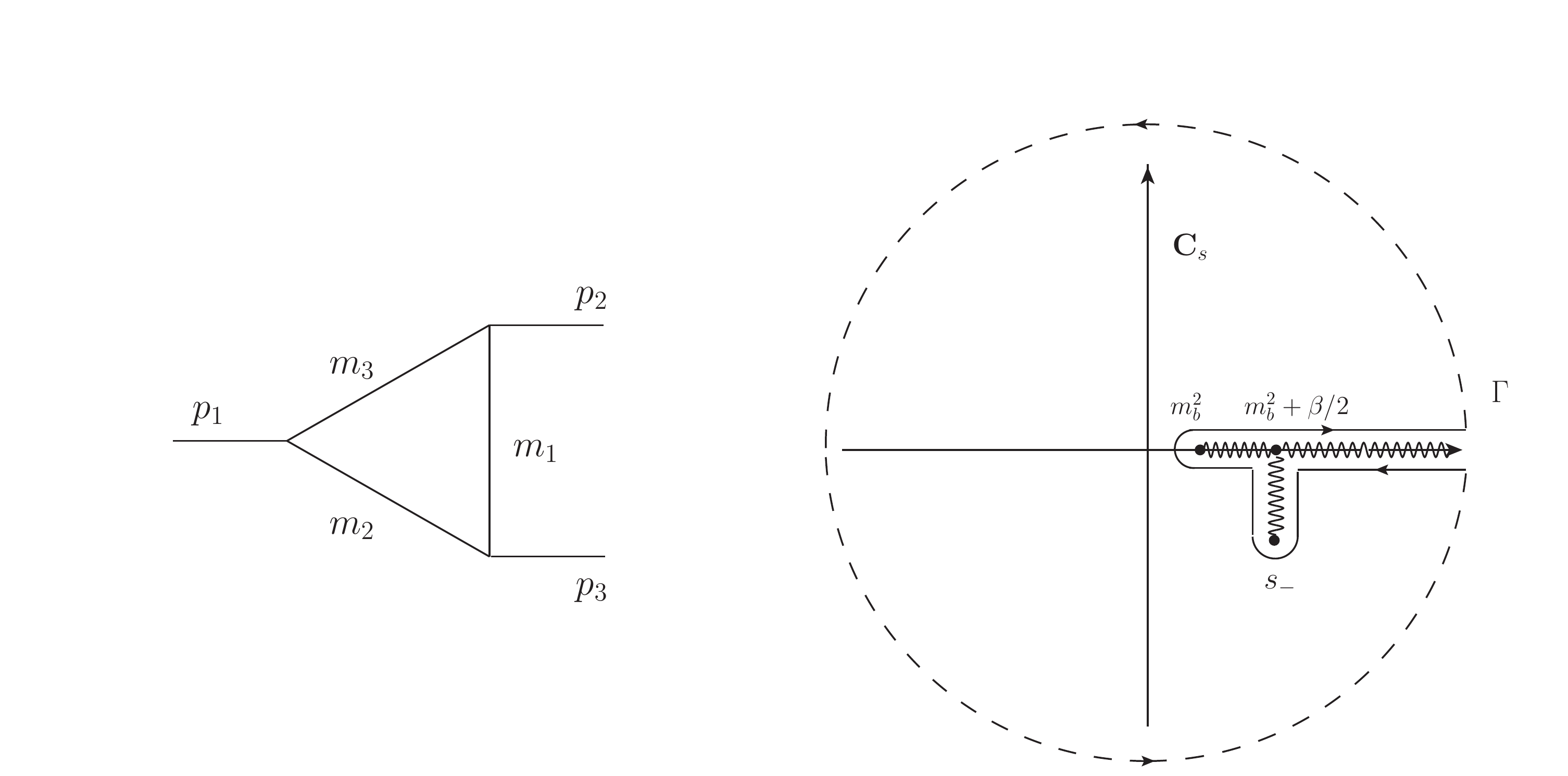}
}
 \caption{\small (left) Triangle graph corresponding to the $C_0(p_1^2,p_2^2,p_3^2,m_2^2,m_3^2,m_1^2)$ PV-function. The conventions are the same as in LoopTools \cite{LoopTools} and Feyncalc 
 \cite{Mertig:1990an}.
 (right) analytic structure of $C_0(s, s - \beta , \alpha  ,0,m_b^2,0)$ in the  $\C_s$-plane; more precisely this corresponds to the physical sheet. The leading Landau singularities of this triangle function are 
 $s_\pm = m_b^2 + \beta/2 \pm \sqrt{(\beta/2)^2 - \alpha m_b^2 }$. As discussed in the text 
 $s_-$ is on the physical sheet whereas $s_+$ is not.}
 \label{fig:triangle}
 \end{figure}

For the important steps of selecting cuts for the dispersion relation
and Borel transformation we refer the reader to the main paper \cite{DLZ12}.
We shall quote here an intermediate expression,
\begin{equation}
\begin{split}
\label{eq:exact}
\matel{\gamma^*(q) V(p)}{\tilde O_8}{\bar B(p_B)}  &= \\
\frac{1}{f_B m_B^2} \frac{1}{2 \pi i}  &\int_{\Gamma \backslash \Gamma_{\rm NP} } ds e^{\frac{m_B^2-s}{M^2}} \Pi^V(q^2, s) \;,
\end{split}
\end{equation}
that relates an on-shell matrix element to an integral over an off-shell correlation function.
$\Gamma$ is a closed path on the physical Riemann sheet that  does not contain any singularities 
such as poles and cuts.  $\Gamma_{\rm NP}$ is of the same type except that it contains the pole of the matrix element.  At this stage the relation is exact but admittedly   rather cryptic. 
In sum rules the analytic structure of the correlation functions is usually such that 
 the singularities are on the real line.  
In the case at hand though it happens that there is an anomalous threshold extending into the complex
plane. Let us explain in more detail: after the reduction to scalar integrals 
the Passarino-Veltman function
 $C_0(s,u (s\!-\!m_B^2),\bar u m_B^2 \!+\!u q^2 ,0,m_b^2,0)$ (where $s=p_B^2$) appears 
 in the expression for $G_\iota^{(s)}(q^2)$, c.f. Fig.~\ref{fig:triangle}.  
The fact that this function has an anomalous threshold extending into the lower half-plane, for $q^2 > 0$ and appropriate momentum fraction $u$, can be seen in various ways.  
First, using the explicit result valid on the real line we can show by uniqueness
of analytic continuation from the real line that there must be singularities in the complex plane 
\cite{DLZ12}.  Second, by setting $m_B^2 < 0$ in the dispersion integral we see that its path 
is deformed into the lower complex plane by analytic continuation of $m_B^2$ to its physical value \cite{DLZ12}.
The analytic structure of this function is depicted in Fig.~\ref{fig:triangle} for a simplified set of variables.
Further to that, the work of K\"all\'{e}n and Wightman \cite{KW58} shows
that anomalous thresholds are present in the corresponding triangle function of the full theory using a minimal number of axioms.
This almost implies that the anomalous thresholds are present in the full theory: the loophole is that the reduction to a scalar object in the non-perturbative case is not as efficient as for the one-loop case;
thus it could in principle be that the contributions cancel, however this is unlikely and in any case not relevant to our discussion.
The important point is that the ${\rm Re}[s_-]$ and thus the rest the anomalous branch cut is well above the $m_B$-pole,
and the anomalous cut is part of what is usually called the continuum contribution in sum rules.  
In the following paragraph we aim to discuss to what extent these anomalous thresholds are surprising 
and what their meaning is in the hope of clarifying to the reader some of our brief argumentation above.

The crucial point is that  the correspondence between matrix elements and correlation functions
is complicated when the number of legs increases.  Let us begin by discussing the simplest case. 
For a two-point function of gauge invariant operators, a dispersion representation 
is in one-to-one correspondence with the insertion of a complete set of states 
as is explicit in the celebrated K\"all\'{e}n-Lehmann representation \cite{KL} and derivations 
thereof. Thus the analytic structure in the complex plane of the four momentum invariant
has a cut and poles on the real line starting from the lowest state in the spectrum.
For correlation functions with three or more fields, there is no such direct relation. 
It seems preferable to think of the analytic structure (singularity structure)  
as a fundamental part of the correlation function rather than as the insertion of a complete set of states. 
For the dispersion relation, which is essentially an application of Cauchy's integral theorem,
it is immaterial whether the singularities are of the normal or anomalous type. 
Normal thresholds are related to unitarity, that is to say to the insertion of a complete set of states.
Anomalous thresholds do not have such an interpretation.
In fact, the anomalous threshold of the triangle graph in perturbation theory
corresponds to the leading Landau singularity which in turn amounts to setting  
all the propagators on the mass shell. 
One out of the two solutions $s_\pm$ of the (leading) Landau equations turns out 
to be the end of the anomalous threshold (Fig.~\ref{fig:triangle}), whereas the other one
is not on the physical sheet. 
We should add that determining which Riemann-sheet  a complex singularity is on is generally a difficult problem.

\section{Results}
\label{sec:results}

We refer the reader to reference \cite{DLZ12} for some explicit analytic results. 
The results of $G_\iota^{(s)}(q^2)$ and  $G_\iota^{(ns)}(q^2)$ are collected in appendices
A and C of that reference. 
Amongst all the possible flavour transitions, only four  are characteristically
different   depending on  whether the initial meson is neutral or charged and
whether it is of the beauty or charm type. Various subparts of this class, at $q^2 =0$, 
are collected in Tab.~\ref{tab:ratios}. 
The ratio of $G_1^{(s)}$ to $T_1(0)$ can be understood as a radiative correction 
and is proportional to $\alpha_s(m_{b(c)})/(4 \pi)$ times other factors of ${\cal O}(1)$.
Further semi-quantitative insight can be gained by considering the heavy quark scaling
of the matrix elements which is $m_b^{-3/2}$ except 
for diagrams $A_{1,2}$ which scale as $m_b^{-5/2}[\ln(m_b) + {\cal O}(1)]$ as discussed in section \ref{sec:QCDfac}.
Projecting out the quark charges the respective 
ratios of $G_1^D(0)$ to $G_1^B(0)$ follow the ratio of heavy quark scaling and 
$\alpha_s(\sqrt{m_c \Lambda_{\rm had}})/\alpha_s(\sqrt{m_b \Lambda_{\rm had}})$ surprisingly well.

\begin{table}[h]
\addtolength{\arraycolsep}{3pt}
\renewcommand{\arraystretch}{1.2}
$$
\begin{array}{ l  |  rr } 
{\rm type}   & B^- \to \rho^- \gamma & \bar B^0 \to \rho^0 \gamma  \\ \hline
G_1^{(s)}(0)  \cdot 10^{2}  &   0.30-0.41i &  0.23+0.21 i   \\
G_1^{(ns)}(0)  \cdot 10^{2}   &  0.90+ 1.3 i   &   0.90+ 1.3 i    \\ 
G_1(0)   \cdot 10^{2} &   1.2+0.89 i & 1.1 + 1.5 i  \\ \hline
 \left| G_1^{(s)}/G_1^{(ns)}  \right|(0)  & 32\% & 20\%  \\
 \left| G_1^{(s)}/ T_1 \right|(0) & 2\%  & 1\%      \\
   \left| G_1/ T_1\right|(0) &  6\%  & 7\%   \\ \hline \hline 
   {\rm type}  & D^+ \to \rho^+ \gamma & D^0 \to \rho^0 \gamma \\ \hline
   G_1^{(s)}(0)  \cdot 10^{2}  &  -1.9+2.6  i &  -7.4-5.2i  \\
G_1^{(ns)}(0)  \cdot 10^{2}   &  -8.5 - 12 i & -8.5 - 12 i  \\ 
G_1(0)   \cdot 10^{2} &   -11 -9.4 i & -16-17 i \\ \hline
 \left| G_1^{(s)}/G_1^{(ns)}  \right|(0)   & 21\%  & 59\% \\
 \left| G_1^{(s)}/ T_1 \right|(0) &  5\% & 13\%     \\
   \left| G_1/ T_1) \right|(0)    &  21\% & 34\%
\end{array}
$$ 
\addtolength{\arraycolsep}{-3pt}
\renewcommand{\arraystretch}{1}
\caption[]{\small  Comparison of various parts of the four characteristic types of  $G_\iota$ matrix elements. 
For the short distance form factor $T_1(0)$  we use  $T_1^B(0) = 0.27$ \cite{BZBtoV} and $T^D_1(0)= 0.7$ \cite{LZ12a} as reference values. 
Recall  $G_1(0) = G_1^{(s)}(0) + G_1^{(ns)}(0)$ and  $G_1^{(s)}(0) = G_1^{(\perp)}(0)$ at our level of twist approximation.} \label{tab:ratios}
\end{table}

\begin{figure}[h]
 \centerline{\includegraphics[width=2.4in]{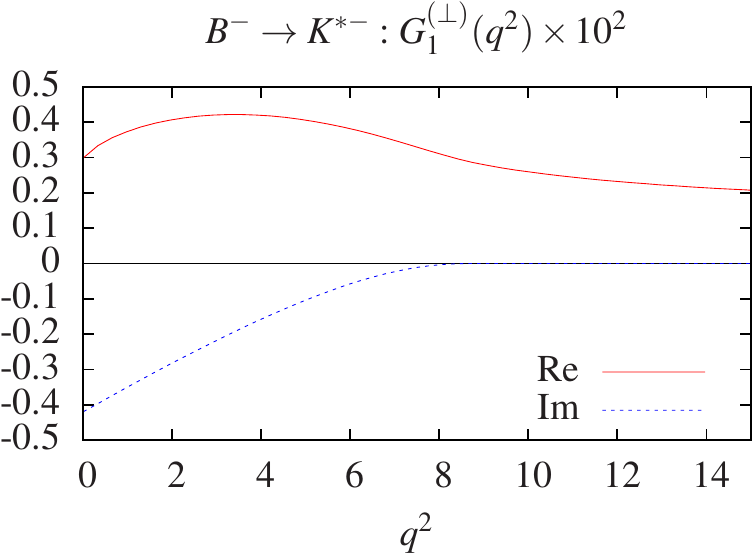} }
 \centerline{ \includegraphics[width=2.4in]{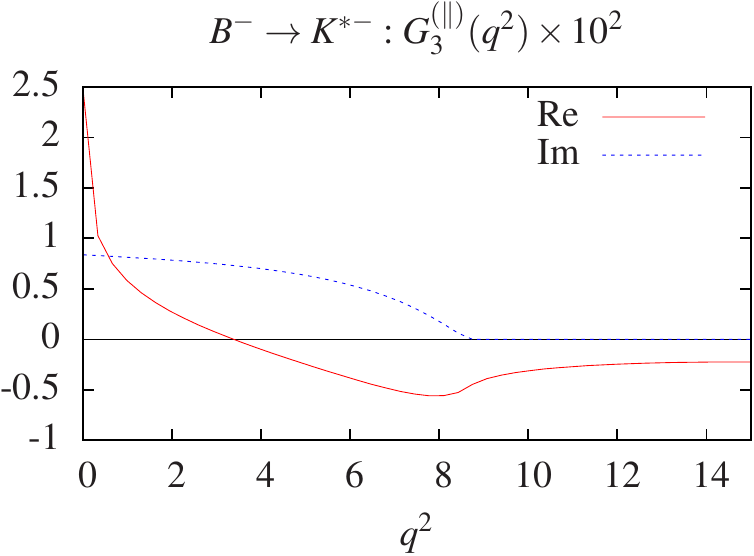}  }
 \caption{See caption of Fig.~\ref{fig:plots}}
 \label{fig:plotsM} 
 \end{figure}

 \begin{figure}[h]
\centerline{ \includegraphics[width=2.4in]{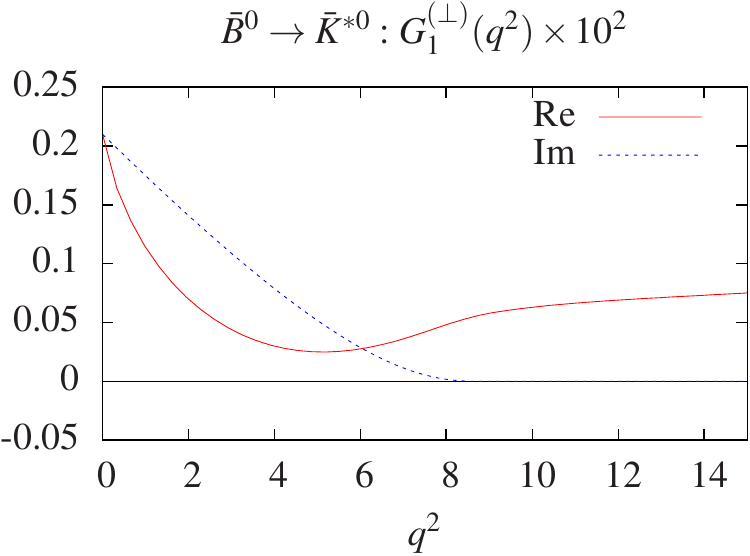} }
\centerline{  \includegraphics[width=2.4in]{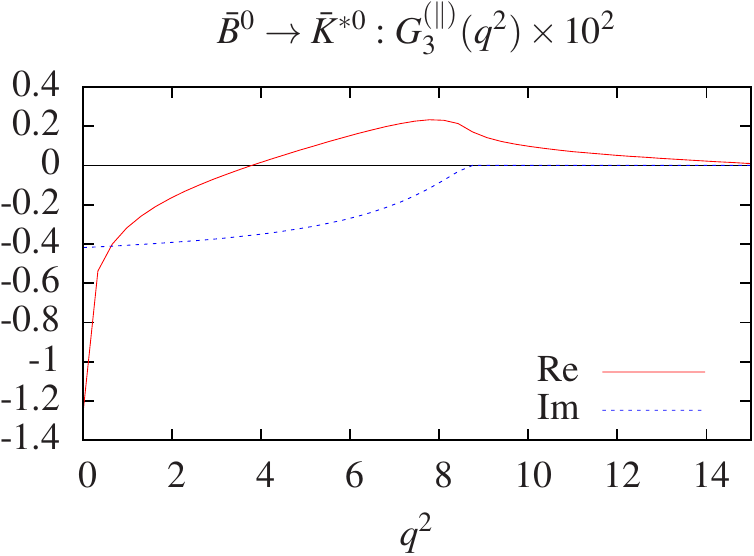}}
 \caption{\small Plots of $G^{(\perp)}_1(q^2)$ and $G_3^{(\parallel)}(q^2)$ for charged (c.f. Fig.~\ref{fig:plotsM}) and uncharged $B$ meson. These plots make it clear why one the $G_\iota(q^2)$-functions are not referred to as form factors as they do not reveal a particular structure but parameterise a multitude of effects amongst which long distance contributions play a definite r\^ole. 
 The results are valid in the region of $q^2$ above $1 \GeV^2$, that is to say outside the resonance region 
 and say $3 \GeV^2$ below $m_b^2$.}
 \label{fig:plots}
 \end{figure}

We consider it worthwhile to quickly mention how the various projections of the DAs contribute 
to the matrix elements and how they are interrelated.  At twist-2 there are seven contributions:
\begin{eqnarray}
\label{eq:Gdec}
G_i^{(s)}  &=&   G_i^{(\perp)}(q^2) + G_i^{(\parallel)}(q^2)   \;,   \nonumber \\[0.1cm]
G_T^{(s)} &=& G_T^{(P)}(q^2)    \;,
\end{eqnarray}
It turns out that (\ref{eq:Gdec}) can be fully reconstructed by knowing 
the three subparts $G_1^{(\perp)}(q^2)$,  $G_3^{(\parallel)}(q^2)$ and $G_T^{(P)}(q^2)$. 
Thus there are four relations:
 $G^{(\parallel)}_1(q^2) = 0$, $G^{(\parallel)}_2(q^2) = 0$, $G^{(\perp)}_2 = (1 - q^2/m_B^2) G^{(\perp)}_3$ and $G^{(\perp)}_2 = (1 - q^2/m_B^2) G^{(\perp)}_1$.  The third relation assures a finite decay width 
 in the limit $m_V^2 \to 0$ (as employed here) \cite{LZ12b}. 
The fourth relation is of the  large energy effective theory (LEET)-type as found for form factors  \cite{LEET98},
whose origin is explained in appendix A of \cite{DLZ} using the third relation and a generic ansatz.
Furthermore, in the ultra-relativistic approximation $m_V^2 \to 0$, the projections  
$G_T^{(P)}(q^2)$ $ G_3^{(\parallel)}(q^2)$  are proportional to each other modulo 
a replacement of the corresponding DA. 
Thus the results can be understood qualitatively from the plots 
of $G_1^{(\perp)}$ and $G_3^{(\parallel)}$ c.f. Fig.~\ref{fig:plotsM},\ref{fig:plots} 
as the DA of $\phi^\parallel(u)$ and $\phi^P(u)$ hardly differ 
in practice. 
The  uncertainties, which are added in quadrature, 
are estimated  to be between  $25\%$ and $35\%$ \cite{DLZ12},  for the $b$ and $c$-transitions respectively, depending on 
the hadronic input. It worthwhile to point out that the uncertainties are higher
for the charm transitions because of the low scale for $\alpha_s$.

\section{Comparison with QCD-factorisation}
\label{sec:QCDfac}

\begin{figure}[h]
 \centerline{\includegraphics[width=3.6in]{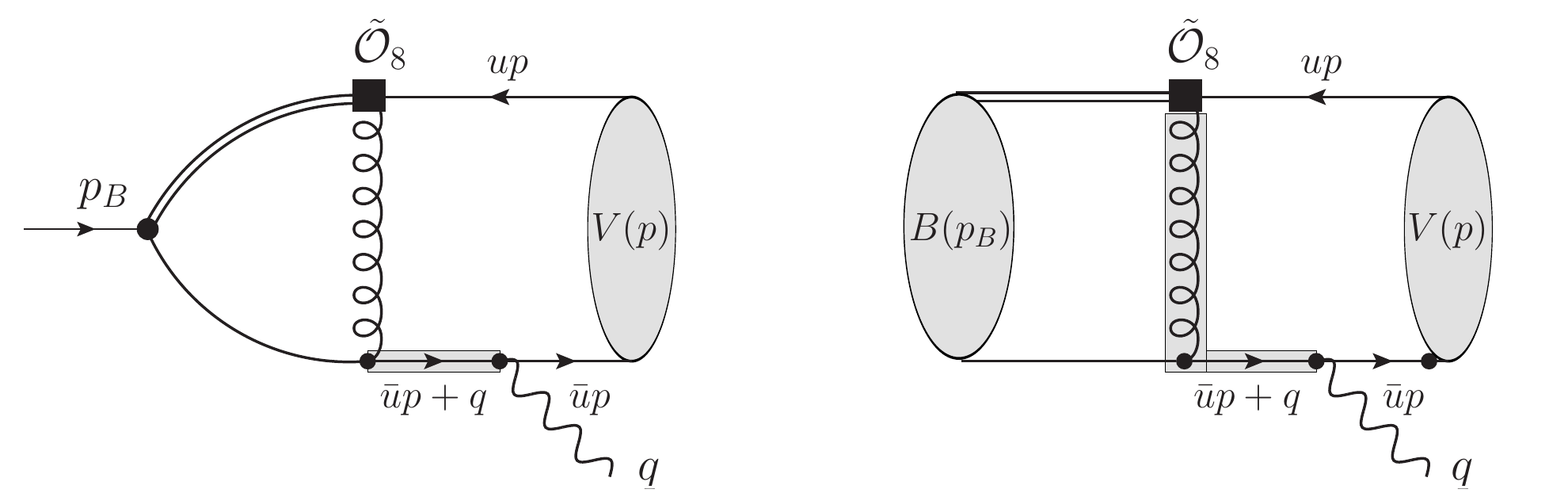}}
 \caption{\small The shaded propagators scale like $1/(\bar u m_B^2)$ in both figures. 
 (left) Diagram used in LCSR. (right) Diagram used in QCDF. 
 Thus $x_\perp^{QCDF} \sim 1/\bar u^2$ and $x_\perp^{LCSR} \sim \ln (\bar u)/\bar u$ at worst, as explained in the text.}
 \label{fig:endpoint}
 \end{figure}

We shall summarise here a few  points of the discussion in chapter 5 of \cite{DLZ12} on 
comparing the contributions of diagrams $A_{1,2}$ of Fig.~\ref{fig:diaA} in the QCDF and LCSR approaches.
These diagrams form a well defined subset as they are isospin dependent. 
Parameterising $G_1(0)$ as
$$
\label{eq:para}
G_1(0) = {\cal O}(m_b^{-5/2})  \underbrace{\int_0^1  \phi_\perp(u) x_\perp(u)}_{\equiv X_\perp}   + {\cal O}(Q_b) \;,
$$
the QCDF \cite{KN01}  and LCSR  \cite{DLZ12} results read
\begin{eqnarray}
\label{eq:xperp}
x_\perp^{QCDF}(u) &=&   \frac{1+\bar u}{3 \bar u^2}  \;, \nonumber \\[0.1cm]
x_\perp^{LCSR}(u)  & = &
\int_{m_b^2}^{s_0} ds \, e^{\frac{m_B^2-s}{M^2}} \rho(s,u)  \;,
\end{eqnarray}
with $\bar u \equiv 1-u$ and
\begin{equation*}
\label{eq:rho0}
\rho(s,u) ={\cal O}(m_b^3) \left[
\frac{\log\left(\frac{\bar us(m_b^2 + m_B^2 - s)}{m_B^2(m_b^2 - us)}\right)}{m_B^2 - \bar us} - \frac{s - m_b^2}{\bar u s m_B^2} \right] \;.
\end{equation*}
The endpoint divergence in QCDF arises as follows: assuming an endpoint behaviour $ \phi_\perp(u) \stackrel{ u \simeq 1}{\to} 6 \bar u u$, which is true at every finite order in the Gegenbauer expansion, it 
is readily seen that $X_\perp^{QCDF}$ is logarithmically divergent (at the endpoint $u =1$).  On a purely technical level this happens
because two propagators behave as $1/ \bar u m_B^2$ c.f. Fig.~\ref{fig:endpoint}. 
In LCSR there is only one  propagator with manifest $1/(\bar u m_B^2)$-behaviour (c.f. Fig.~\ref{fig:endpoint})
and there is no such term hidden in the loop as it would correspond to a  
power IR-divergence, and it is known that in 
four dimensions IR-singularities, whether they are soft or collinear, are at worst of logarithmic nature, e.g. \cite{Muta}.
Thus the worst behaviour that we can get is $x^{LCSR}_\perp \sim \ln \bar u  /\bar u $ which is integrable, i.e.
does not show an endpoint-divergence.
Inspection or evaluation Eq.~(\ref{eq:xperp}) gives an even 
milder behaviour with $x^{LCSR}_\perp \sim \ln  \bar u$ for which we see no particular reason.

\begin{figure}[h]
 \centerline{\includegraphics[width=3in]{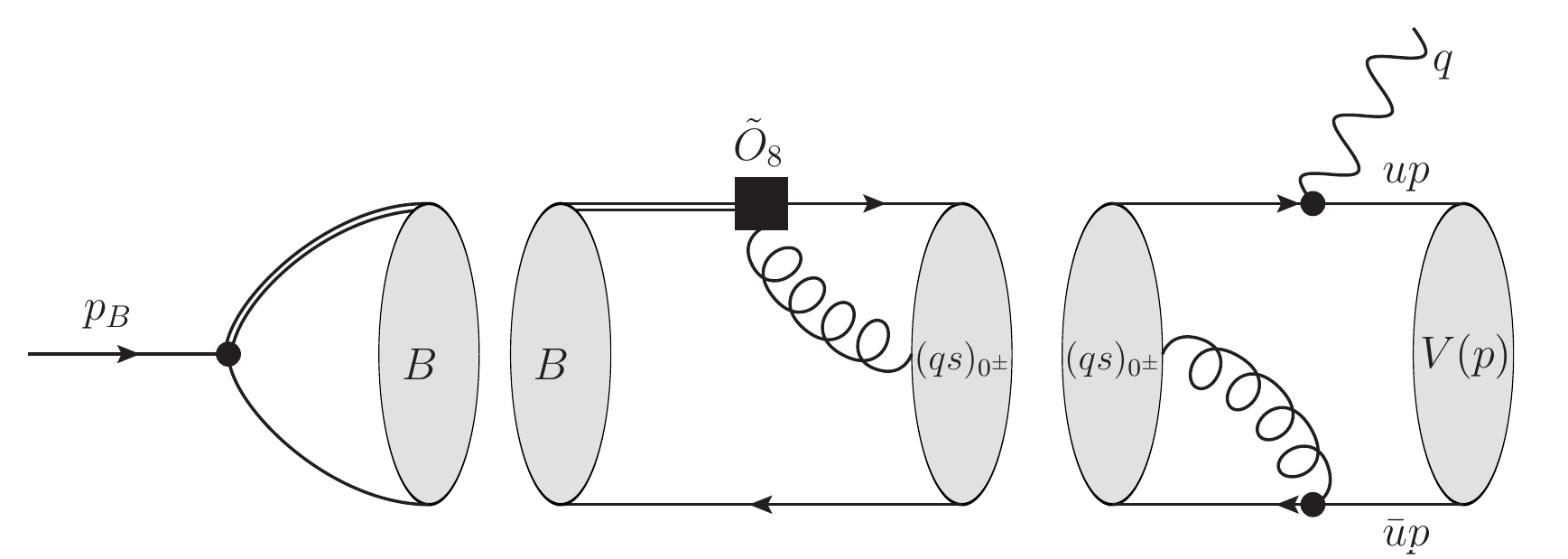}}
 \caption{\small (left) Hadronic interpretation of the $3$-particle cut 
 in  in terms of a long-distance hadronic process. The state $(\bar q s)_{0^\pm}$ is any state, single or multiparticle,  of spin zero with 
 $\bar q$ and $s$ quantum numbers.
 The latter is a source 
 for the strong (CP-even) phase that we obtain for the  $G_\iota(q^2)$-functions.}
 \label{fig:phase_P}
 \end{figure}

The comparison between QCDF and LCSR can be sharpened if the heavy quark scaling
$m_B \to  m_b + \bar\Lambda$,  
$s_0 \to  m_b^2 + 2m_b\omega_0$ and 
$M^2 \to  2m_b\tau$, as suggested in \cite{CZ90}, is applied:
\begin{equation}
X_\perp^{LCSR} \sim \ln (m_b/2 \omega_0) + {\cal O}(1) + {\cal O}\left( \frac{\Lambda_{\rm QCD}}{m_b} \right)  \;.
\end{equation}
The logarithmic term indicates that this expression is not expandable in inverse powers of the
heavy quark mass. 
In fact, the expansion of the density around $m_b=\infty$ reveals that  $\rho(s,u) \sim 1/(m_b^n \bar u^{n+1})$
in our results. Are the LCSR and the QCDF to be seen on an equal footing 
when the leading heavy quark scaling behaviour is considered? The answer must be no
as the former has a sizeable imaginary part whereas the latter is real (at leading order in $\alpha_s$) 
\cite{DLZ12}.  This difference is due to the fact that LCSR contain additional contributions which are not 
present in the QCDF computation.  An interpretation of the complex phase as a long distance phenomenon 
is given in the caption of Fig.~\ref{fig:phase_P}. 

In regards to the discussion above it is seems worthwhile to point out that even though QCDF and LCSR 
are both based on LC-expansions in terms of hard kernels and DAs they differ on a conceptual level. 
QCDF computes physical processes in a direct way and is in that sense very transparent. 
LCSR is of an indirect nature as correlation functions are computed in which the matrix element in question appears as a residue of a pole.  
The matrix element is then extracted by manipulating the sum rule (Borel transformation) 
and considering appropriate kinematical limits. In this sense LCSR is akin to lattice-QCD extraction of matrix elements.

\section{Conclusions}
\label{sec:conclusions}

In this write up we have reported on the computation of matrix elements of heavy to light meson transitions induced by the ${\cal O}_8$ operator using LCSR.
We focused on the appearance of a complex anomalous threshold and on comparing our results with those previously computed in QCDF. 
A characteristic feature of the results is a large strong phase
reflecting the  long distance physics contained in these matrix elements.

\section*{Acknowledgements}
RZ is grateful for the support of an advanced STFC fellowship and to the organisers of the
4th Capri Workshop on Flavour Physics and to many colleagues for inspiring discussions. 
Computer manipulation and numerics are performed with the aid of LoopTools \cite{LoopTools} 
and FeynCalc \cite{Mertig:1990an}.




\nocite{*}


\end{document}